\documentclass[amsmath, a4paper, prl, twocolumn, showpacs]{revtex4}

\usepackage{amssymb}
\usepackage{amsmath}
\usepackage{amsthm}
\usepackage{epsfig}
\usepackage{color}
\usepackage{graphics, graphicx}
\usepackage{psfrag}
\usepackage{mathcomp}

\begin{document}

\author{Michiel Snoek}
\author{Walter Hofstetter}
\affiliation{Institut f\"ur Theoretische Physik, Johann Wolfgang Goethe-Universit\"at, 60438 Frankfurt/Main, Germany}

\title{Two-Dimensional Dynamics of Ultracold Atoms in Optical Lattices}

\begin{abstract}
We analyze the dynamics of ultracold atoms in optical lattices induced by a sudden shift of the underlying harmonic trapping potential. In order to study the effect of strong interactions, dimensionality and lattice topology on transport properties,  we consider bosonic atoms with arbitrarily strong repulsive interactions, on a two-dimensional square lattice and a hexagonal lattice. On the square lattice we find insulating behavior for weakly interacting atoms and slow relaxation for strong interactions, even when a Mott plateau is present, which in one dimension blocks the dynamics. On the hexagonal lattice the center of mass relaxes to the new equilibrium for any interaction strength.
\end{abstract}

\maketitle

\begin{figure}
\includegraphics[width=8cm]{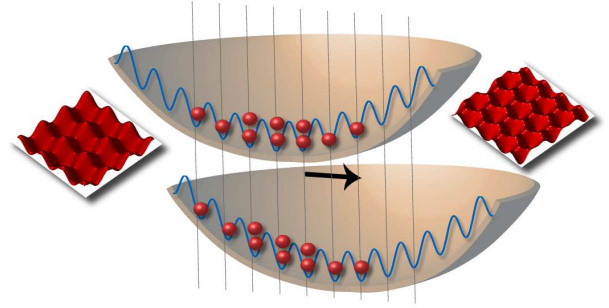}
\caption{(Color online) Schematic of the setup: the harmonic potential is suddenly shifted, bringing the system far out of equilibrium. In the insets the two lattice geometries we study are shown: the square lattice (left) and the hexagonal lattice (right).    
  }
\label{schematic}
\end{figure}

The possibility to confine ultracold atoms in optical lattices has opened up a new research area, where interacting quantum many-body systems consisting of bosons \cite{Jaksch98, Greiner02} and fermions \cite{Hofstetter02, Kohl05} can be studied with unprecedented high precision and tunability \cite{Bloch07}. A new and exciting development is to study the dynamics and out-of-equilibrium behavior of those systems, see e.g. \cite{Morsch06}, which can also be used as an experimental probing technique \cite{Stoferle04, Kollath06}. In particular, it is possible to bring the system far out of equilibrium by performing an instantaneous shift in the underlying harmonic trapping potential. In this way particle transport can be investigated. This procedure is schema\-tically illustrated in Fig. \ref{schematic}. Experiments in this setup with a three-dimensional Bose-Einstein condensate subject to a one-dimensional optical lattice revealed a transition from coherent to dissipative behavior \cite{Cataliotti01, Burger01, Cataliotti03, Tuchman06}.  For a truly one-dimensional gas, the impact of the optical lattice was found to be even stronger: already for small lattice depths dissipative motion was observed \cite{Fertig05, Henderson06}, which has been theoretically investigated with a variety of methods \cite{Gea06, Rigol05, Polkovnikov04, Ruostekoski05, Pupillo06, Rey05}. A three-dimensional fermionic cloud subject to a one-dimensional optical lattice showed strongly suppressed center of mass motion \cite{Mudugno03, Pezze04}. Related theoretical \cite{Altman05, Polkovnikov05} and experimental \cite{Mun07} work on a Bose gas in a moving optical lattice showed a momentum-dependent breakdown of superfluid motion.

However, both experimental and theoretical research in this direction has up to now \cite{Zuerich} been restricted to a one-dimensional optical lattice. In this Letter we analyze the behavior of repulsively interacting bosons in two dimensions, where geometrical considerations play a profound role. The influence of the lattice dimensionality is most clearly seen when a Mott plateau is formed. Unlike in one spatial dimension, where this leads to insulating behavior \cite{Rigol05}, we find that in two dimensions there is always relaxation of the center of mass to the new equilibrium position.  This behavior even persists in the limit of hard-core bosons. 
Moreover, in two dimensions, one can choose different lattice geometries, which severely influences the dynamics. Experimentally this is possible by varying the angle between the laser beams which make up the optical lattice. Here we compare the behavior on the square lattice to the hexagonal lattice and find remarkably different behavior. This originates from the single particle dynamics, which on the square lattice can be described in terms of Bloch oscillations. Those are absent on the hexagonal lattice, where the atoms move along the equipotential lines. Therefore, strongly interacting bosons on the hexagonal lattice show a much quicker relaxation of the center of mass than on the square lattice. In order to study bosonic atoms with arbitrarily strong repulsive interactions we apply the Gutzwiller mean-field theory. This includes the limit of infinite interaction strength (hard-core bosons).

For a deep optical lattice and moderate filling, the bosons can be described by the single band Bose-Hubbard Hamiltonian
\begin{equation*}
\mathcal{H} = - J \sum_{\langle i j \rangle} b_i^\dagger b_j + \sum_i \left\{\frac{U}{2} \hat n_i (\hat n_i - 1) + \left(V(i, t)  - \mu \right) \hat n(i) \right\},
\end{equation*}
where $b_i$ is the bosonic annihilation operator at site $i$, $\mu$ is the chemical potential,
$J$ is the hopping amplitude and $U$ is the on-site repulsion. $J$ and $U$ can be expressed in terms of the atomic interparticle scattering length $a$, mass $m$ and laser wavelength and intensity \cite{Jaksch98}. We will use them as effective parameters.  $V(i, t)$ is the underlying harmonic potential which we take equal to:
\begin{equation}
V(i ,t ) = V_0 |{\bf x}_i - {\bf x}_0(t)|^2.
\end{equation}
In the following we set $J = 1$ and use $U=U/J$, $\mu= \mu/J$ and $V_0 = V_0/J$ as dimensionless parameters. Time is expressed in units of $\hbar/J$. The shift $A$ in the harmonic potential is expressed in terms of the lattice spacing $a$.

\begin{figure}
\includegraphics[scale=.28]{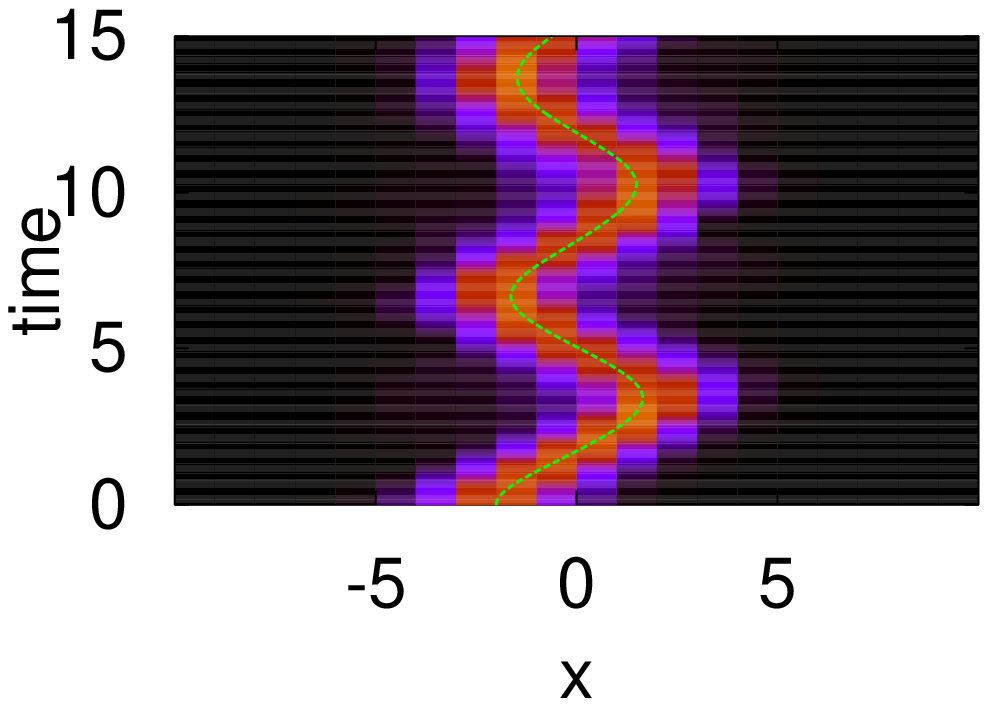}
\hspace{-1.3cm}
\includegraphics[scale=.28]{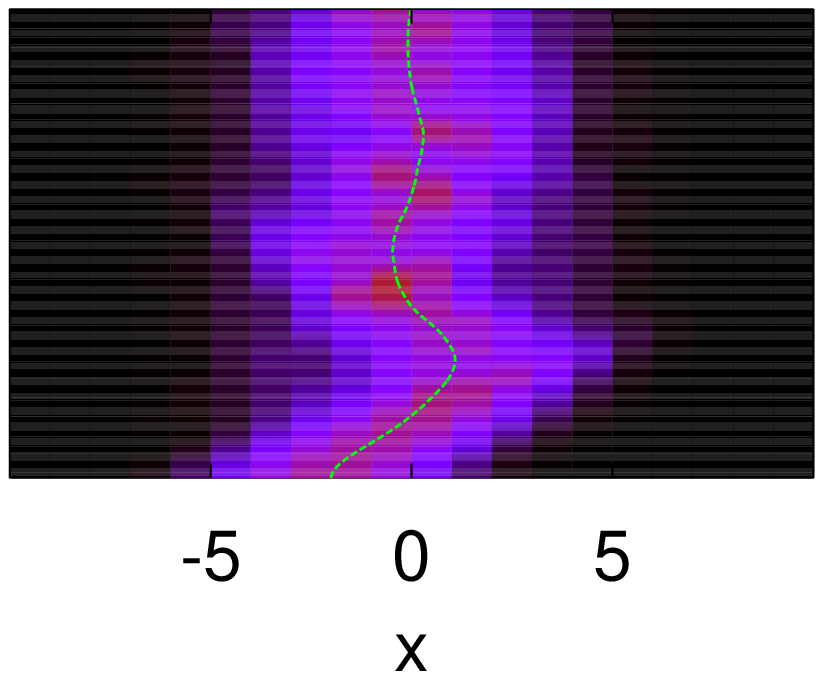}
\hspace{-1.3cm}
\includegraphics[scale=.28]{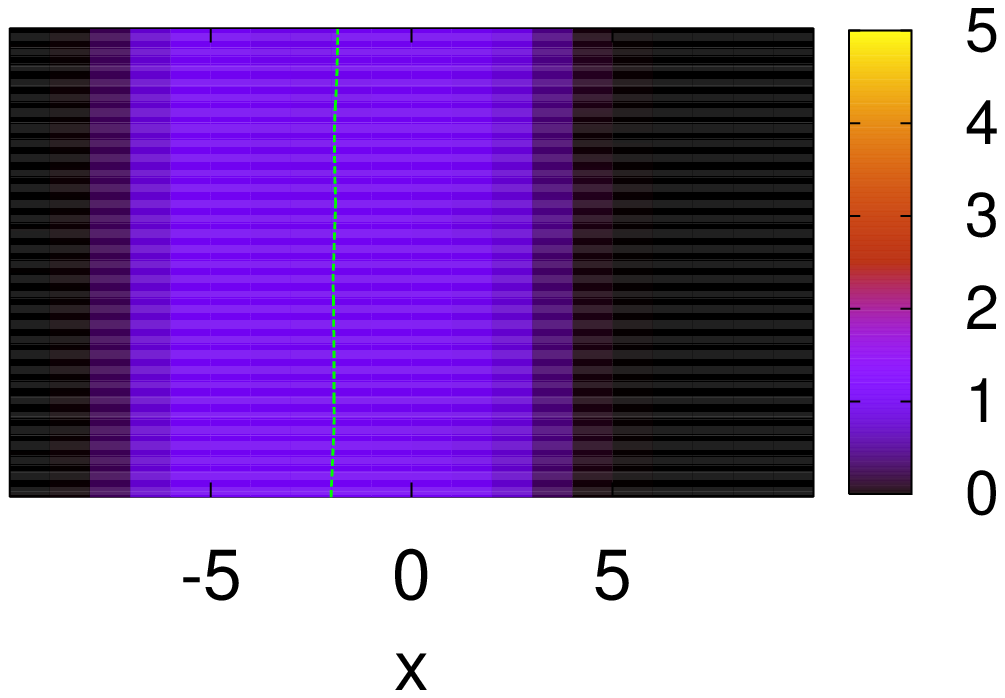}
\vspace{-.5cm}
\caption{(Color online) Density distribution (color coding) and center of mass (green line) as a function of time for a one-dimensional system. Parameters are chosen as $V_0=0.3$, $A=2$, $N=20$ and from left to right $U=0.2$, $U=2$ and $U=20$. For the latter case a Mott plateau forms, which completely blocks the relaxation of the center of mass. }
\label{1dmott} 
\end{figure}

For weakly interacting bosons, a Bose-Einstein condensate forms, whose  dynamics can be described by the Gross-Pitaevskii equation. 
A complementary approach, which is also valid for strongly interacting bosons, is the time-dependent Gutzwiller technique \cite{Jaksch02} where the coupling between the lattice sites is treated in a mean-field approximation. 
For inhomogeneous systems, this procedure has to be carried out in a space-resolved version, where with each site a separate order parameter is associated.
The total many-body wavefunction is within this approximation given as
$
| \Psi \rangle = \prod_i \sum_{n=0}^{\infty} f_n^i \frac{( b_i^\dagger)^n } {\sqrt{n!}} |0 \rangle.
$ 
In practice, the infinite sum over the particle numbers $n$ is replaced by a finite sum, by introducting a cut-off $N_c$ depending on the strength of the interaction and the local density. 
The dynamics is governed by the set of coupled differential equations \cite{Jaksch02}
\begin{eqnarray}
i \dot f_n^i &=& - \sum_{ \langle i j  \rangle }  \left( \sqrt{n+1} \; \Phi_j^* \; f_{n+1}^i + \sqrt{n} \; \Phi_j \; f_{n-1}^i \right) \nonumber \\ 
&& +  \left( \frac{U}{2} n (n-1) + V(i,t) - \mu \right) f_n^i,
\end{eqnarray}
where $\Phi_i = \langle b_i \rangle = \sum_n \sqrt{n} (f_{n-1}^i)^* f_n^i$.
This Gutzwiller approximation is a highly efficient method for studying dynamics in higher dimensional lattices. It conserves energy exactly and particle number with a very good accuracy. The latter, however, is only true if the sites are sequentially updated; a parallel update of all sites together violates particle number conservation. The validity of the Gutzwiller approximation is further justified by the fact that for small interactions it incorporates the Gross-Pitaevskii dynamics \cite{Jaksch02}. 
Also the limit of strong interactions in one dimension is correctly reproduced by the Gutzwiller approximation (see Fig. \ref{1dmott}): in that case the dynamics of the center of mass is completely blocked \cite{Rigol05}. 
The Gutzwiller approximation is naturally restricted to zero temperature, since it neglects phase fluctuations. We therefore only consider $T=0$ here.

\begin{figure}
\includegraphics[scale=.33]{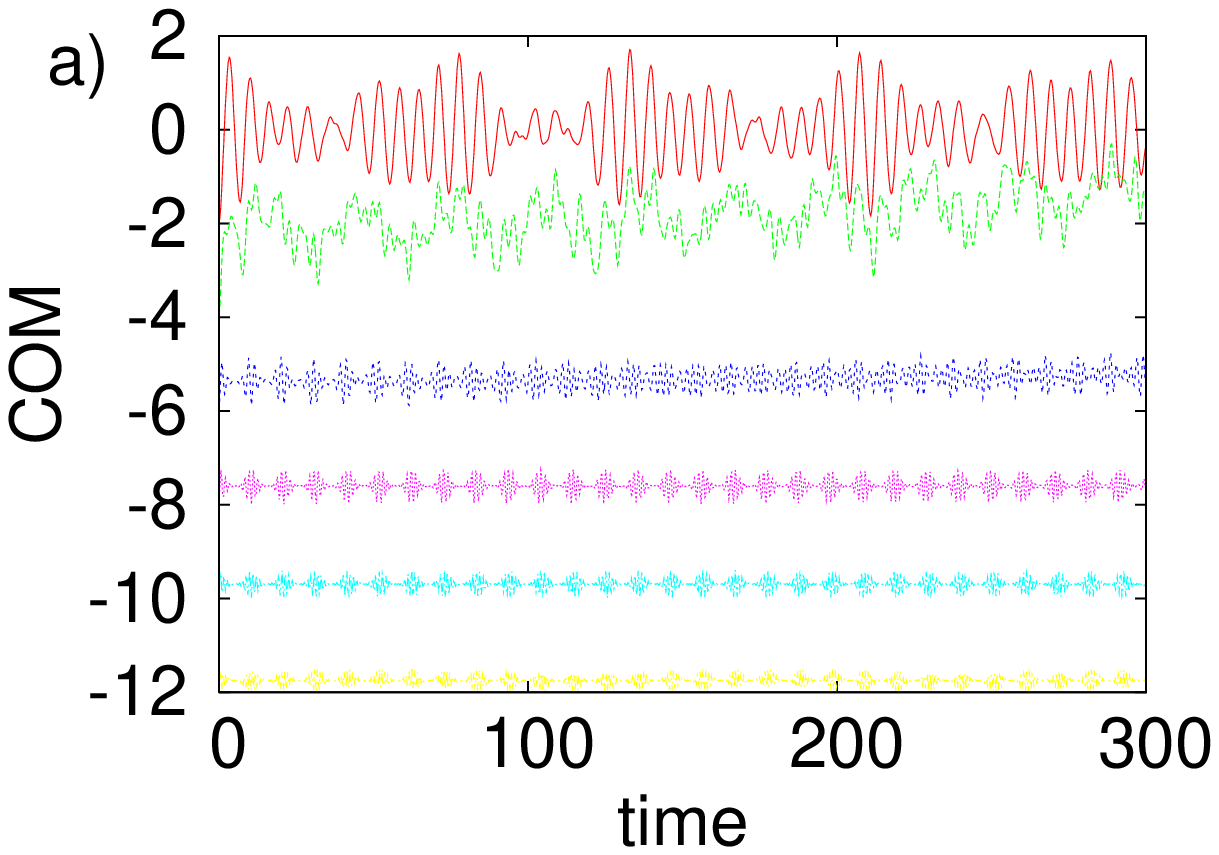}
\includegraphics[scale=.33]{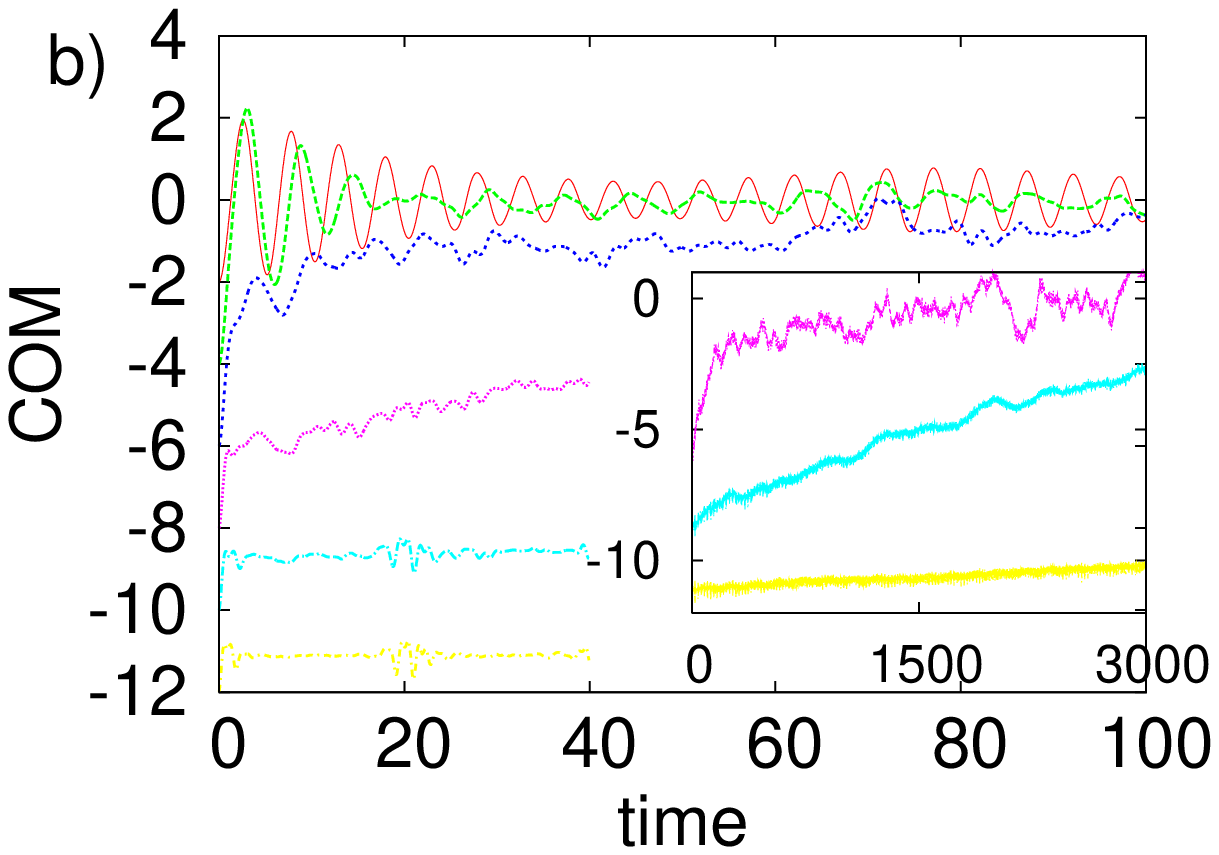}
\includegraphics[scale=.33]{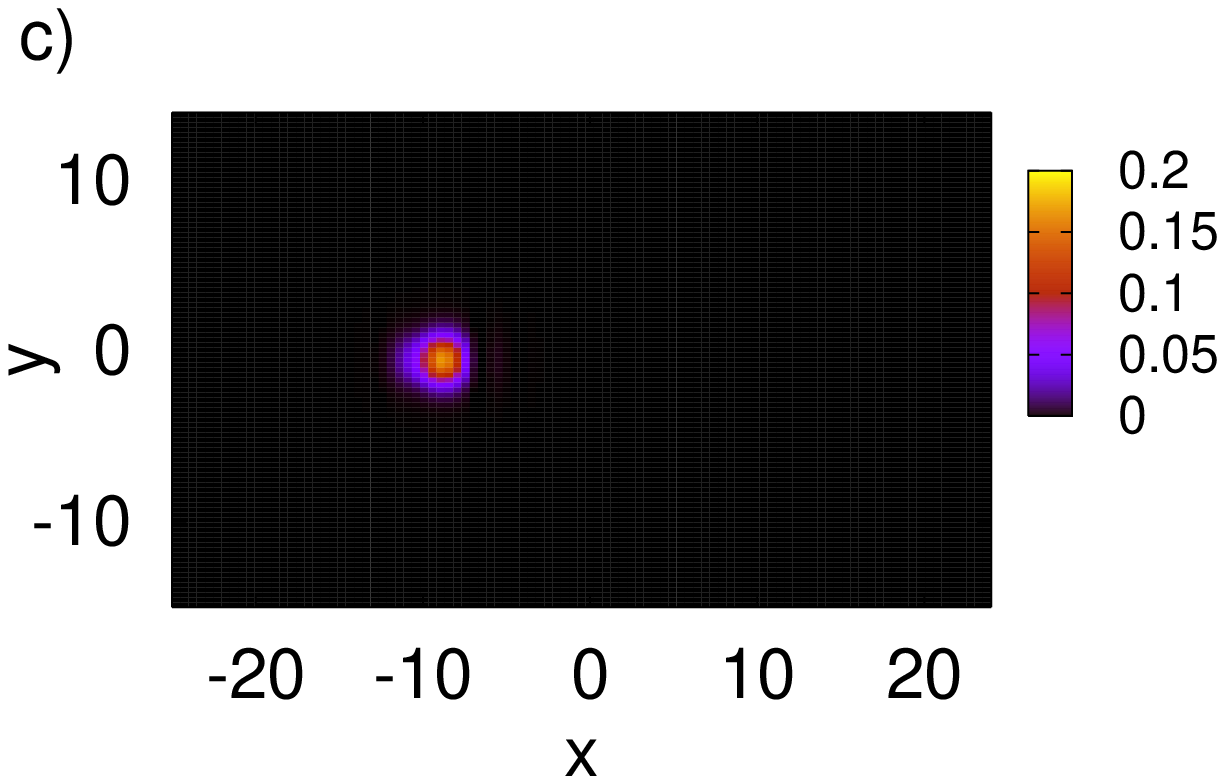}
\includegraphics[scale=.33]{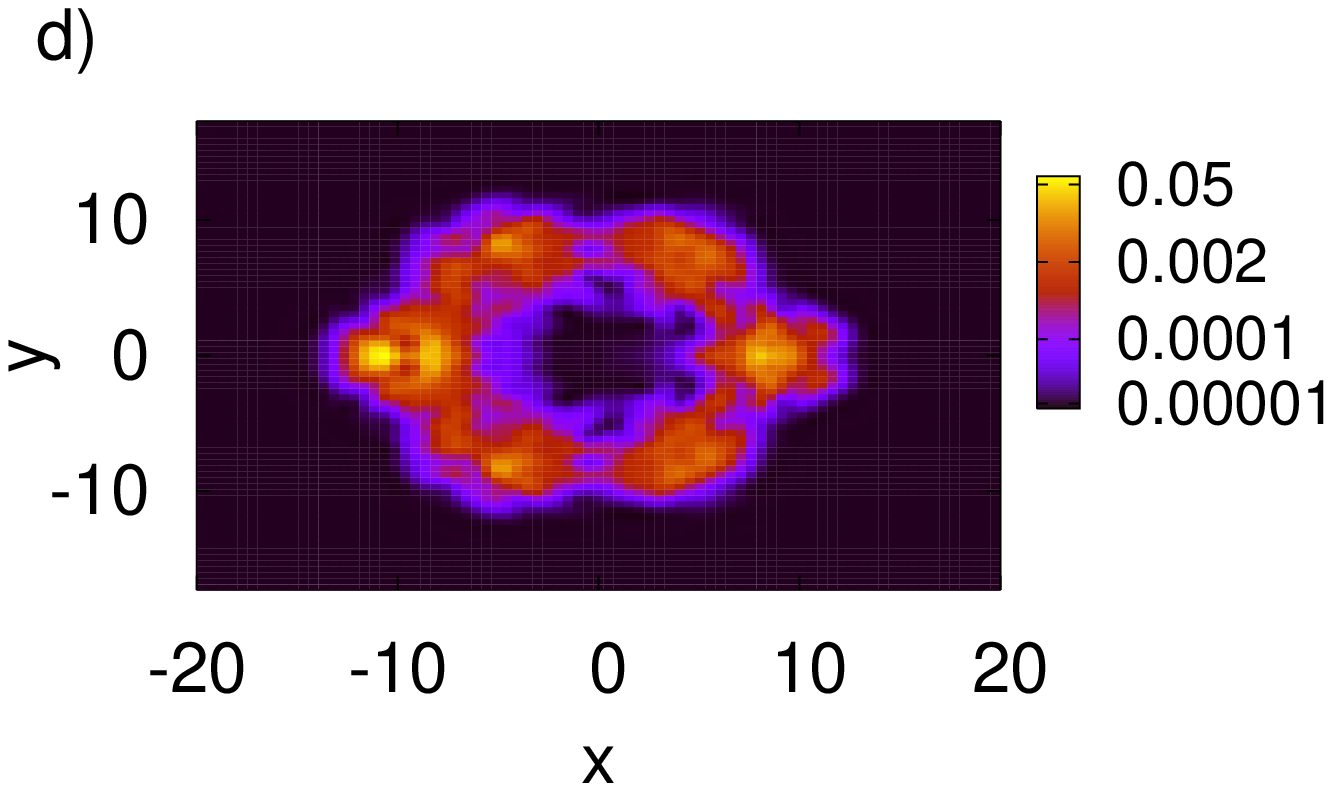}
\vspace{-.5cm}
\caption{(Color online) Center of mass (COM) dynamics for non-interacting bosons after an instantaneous shift in the harmonic potential on a two-dimensional square lattice (left column) and a  hexagonal lattice (right column) obtained with exact diagonalization. For all plots $V_0=0.3$. Shown are shifts $A=2$ (red), $A=4$ (green), $A=6$ (blue), $A=8$ (purple), $A=10$ (light-blue) and $A=12$ (yellow). In b) the long-term dynamics is depicted in the inset. 
In the two lower panels the density distribution of non-interacting bosons is plotted c) for the square lattice, ($t/J=300$)  and d) for the hexagonal lattice ($t/J=3000$) for a shift of ten lattice sites. Note that for the hexagonal lattice the color coding is on a logarithmic scale.
}
\label{2d}
\end{figure}
We first investigate the influence of the lattice topology on the single particle dynamics, which will reflect itself in the behavior of weakly interacting bosons.  
For small shifts, the bosons perform dipole oscillations around the shifted center. Interactions between the bosons lead to damping that increases with the strength of the interaction. This behavior is limited to small shifts, for which the shifted wavefunction still has an overlap with the groundstate-wavefunction in the shifted potential. Alternatively, one can argue that for larger shifts the potential energy introduced into the system by performing the shift cannot totally be converted into kinetic energy, which is restricted because of the single-band description.  
For larger shifts the lattice structure becomes very important, which is most clearly seen in the limit of non-interacting bosons (see Fig. \ref{2d}). If $U=0$, the Hamiltonian on the \emph{square lattice} is the sum of two commuting one-dimensional Hamiltonians: $\mathcal{H}_{U=0} = \mathcal{H}_x + \mathcal{H}_y$ with $\lbrack \mathcal{H}_x, \mathcal{H}_y \rbrack = 0$. This implies that the dynamics is effectively one-dimensional and the single particle eigenstates are products of the one-dimensional eigenstates, which are highly localized \cite{Hooley04}. As a result, the atoms perform Bloch oscillations around the shifted position instead of dipole oscillations after  a large shift \cite{Ponomarev05} as shown in Fig. \ref{2d}a). Because of the non-linearity of the harmonic potential, the Bloch oscillations contain multiple frequencies. Because the eigenstates are localized, the wavepacket remains localized as well (Fig. \ref{2d}c).
The Bloch oscillations persist for small interactions between the bosons, which however cause the Bloch oscillations to be damped, due to dephasing \cite{Buchleitner03}. For stronger interactions the Bloch oscillations disappear. Instead, the center of mass shows dissipative dynamics. This is possible, because in this case the potential energy can be converted into interaction energy, and the center of mass can relax to zero. 

The description in terms of Bloch oscillations does not apply to other lattice structures, where the Hamiltonian cannot be decomposed into two commuting one-dimensional parts. In particular, we have investigated the behavior of free and weakly interacting bosons on a \emph{hexagonal lattice}. In this case we find for arbitrarily large shifts a relaxation of the center of mass to the minimum of the harmonic potential  (Fig. \ref{2d}b)). Inspection of the density profile shows that this occurs, because on the hexagonal lattice the atoms move along the equipotential lines.  The reason for this behavior is, that on the hexagonal lattice plus harmonic potential the single-particle eigenstates form a ring-like structure. Therefore, the bosons perform a ring-like expansion after a large shift and although the center of mass relaxes to the new equilibrium, the atoms actually never do. 

It is worth noticing that spinless fermions behave in the same way as weakly interacting bosons. On the square lattice they perform Bloch oscillations after a large shift \cite{Pezze04}, whereas on the hexagonal lattice the center of mass relaxes to zero, whereas the density forms a ring-like structure. This means that, especially on the square lattice, spinless fermions behave qualitatively different from hard-core bosons. The latter show relaxation of the center of mass to the new equilibrium, as we will demonstrate later. This is unlike the one-dimensional situation, where hard-core bosons behave as free fermions.  

\begin{figure}
\includegraphics[scale=.33]{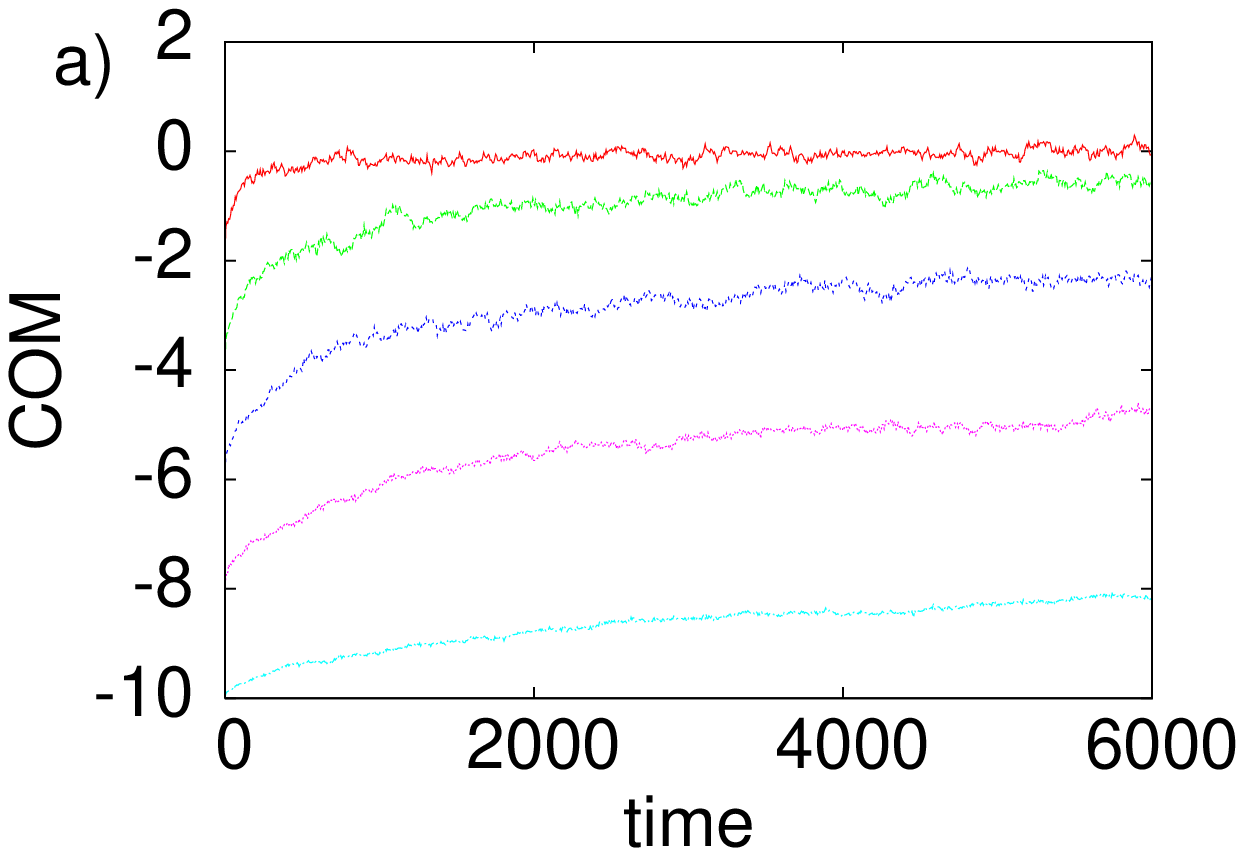}
\includegraphics[scale=.33]{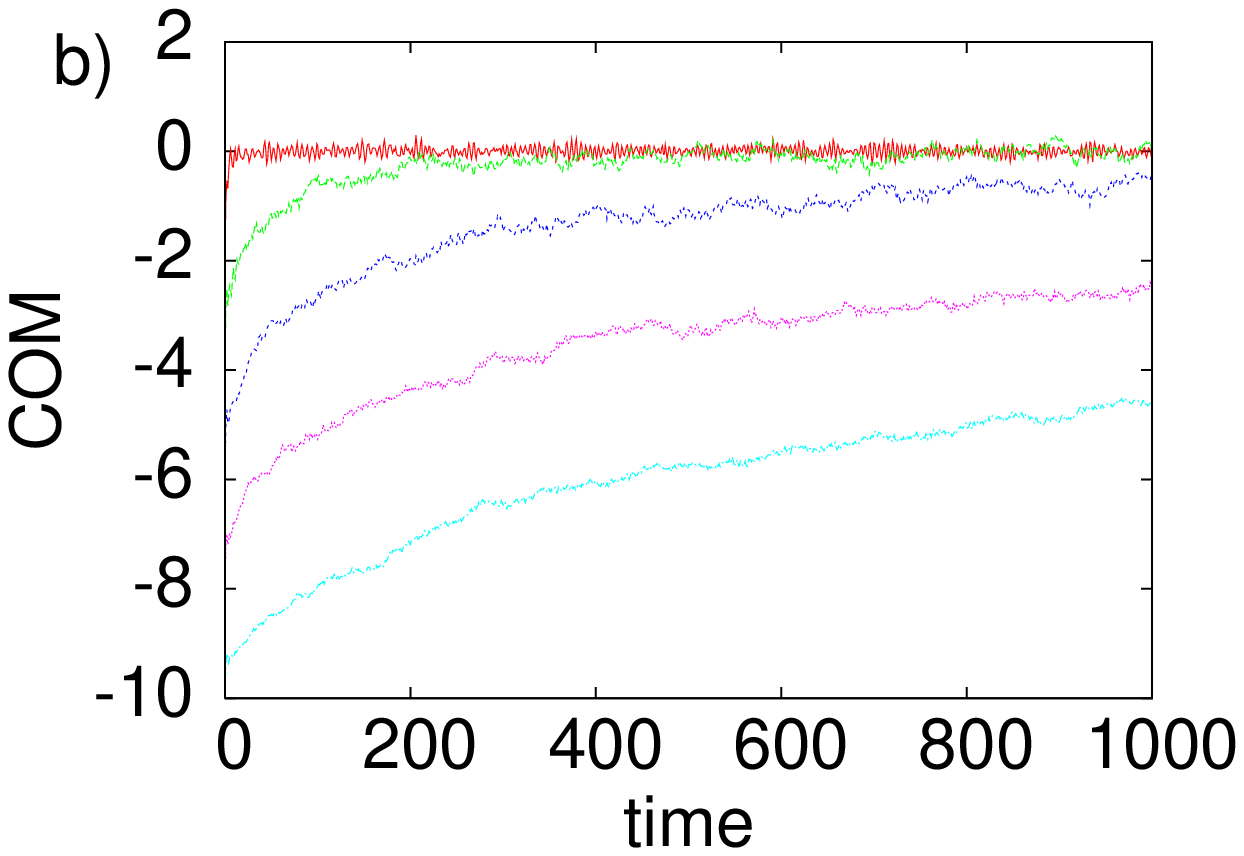}
\includegraphics[scale=.33]{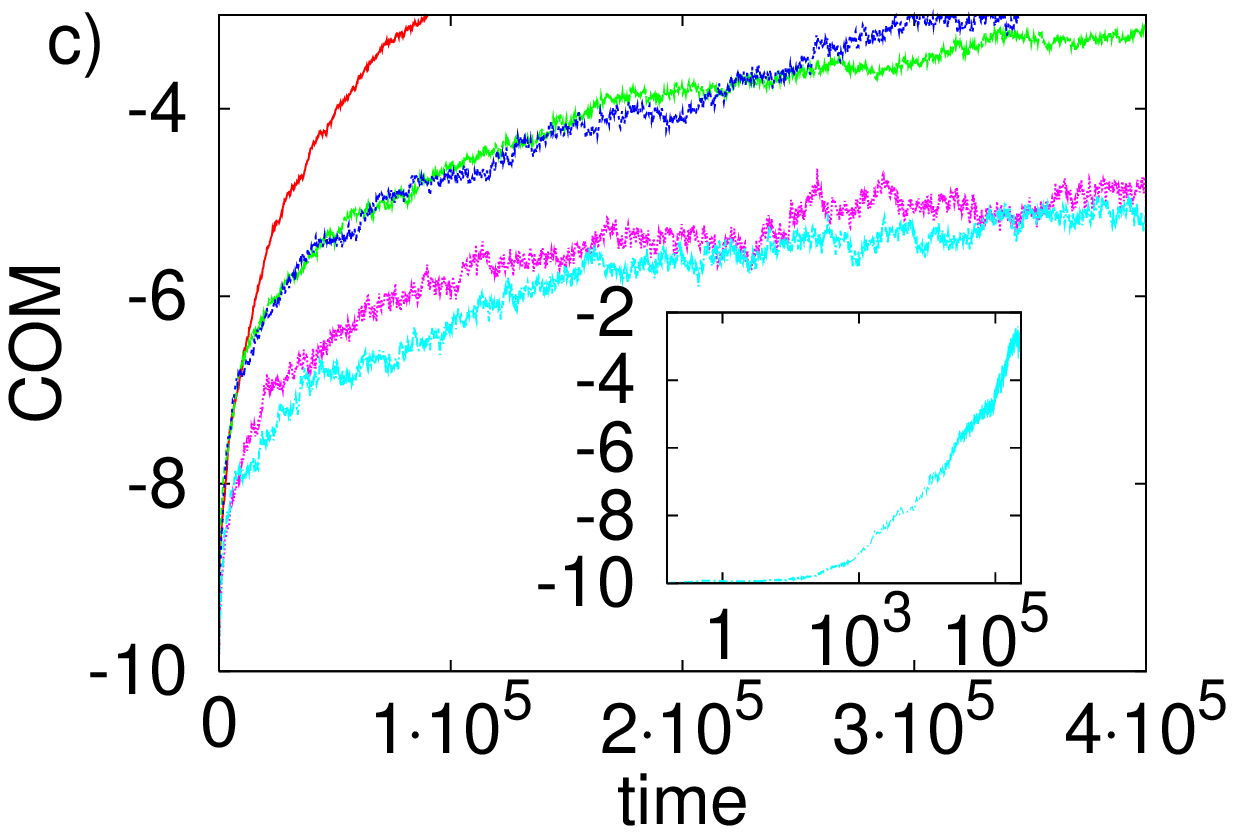}
\includegraphics[scale=.33]{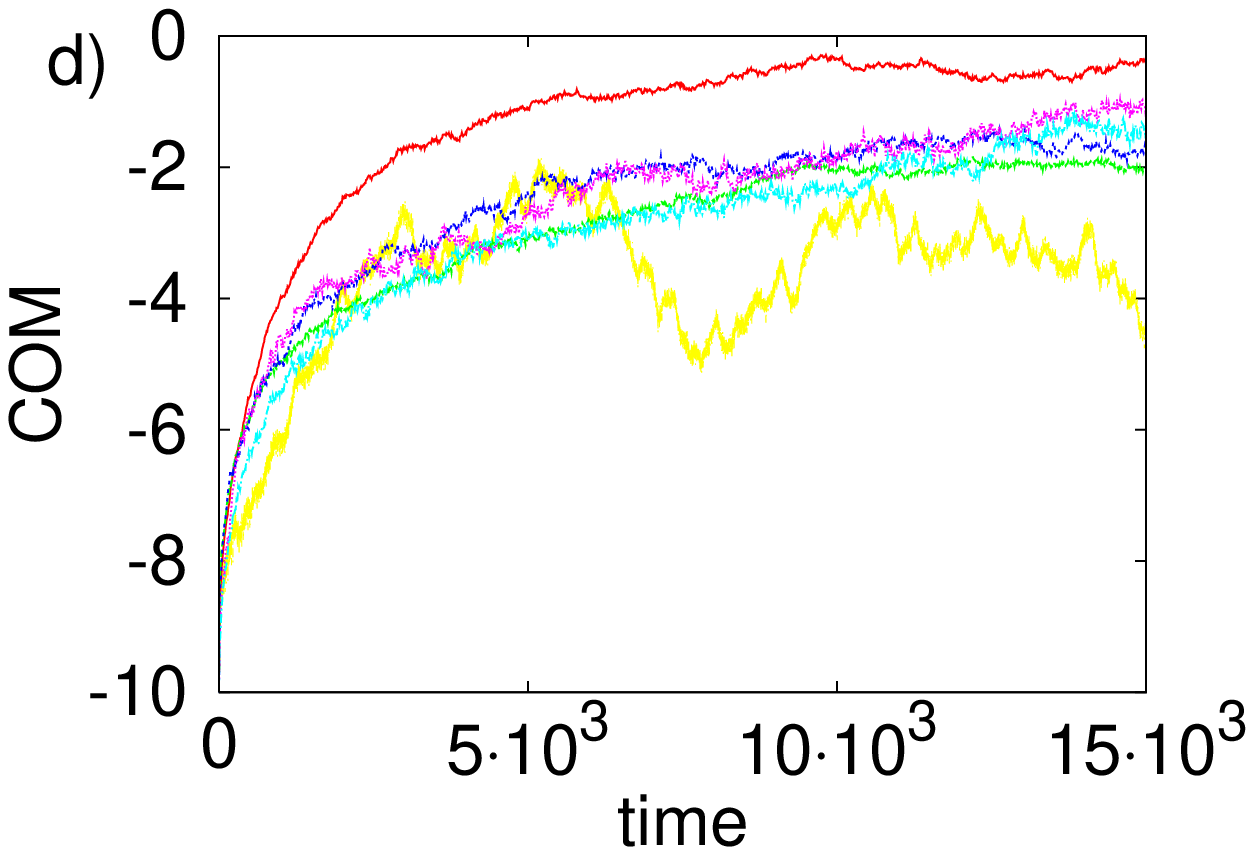}
\vspace{-.2cm}
\caption{
(Color online) Center of mass dynamics as a function of time for strongly interacting bosons on the square lattice (left column) and hexagonal lattice (right column) for \mbox{$N=100$} and $V_0=0.3$. The upper plots are for $U=40$, where a Mott plateau is present before the shift. 
Displayed is the dynamics after a shift of $A=2$ (red), $A=4$ (green), $A=6$ (blue), $A=8$ (purple) and $A=10$ (light-blue). The lower panels show the long-time dynamics for a large shift of $A=10$ for different interaction strengths $U=2$ (red), $U=10$ (green), $U=20$ (blue), $U=40$ (purple) and $U=\infty$ (light-blue). In the inset of c) the long-time dynamics for the center of mass of hard-core bosons is shown on a logarithmic time-scale. In d) the $U=0$-result is depicted in yellow for comparison. 
} 
\label{2dcom}
\end{figure}

\begin{figure}
\includegraphics[scale=.33]{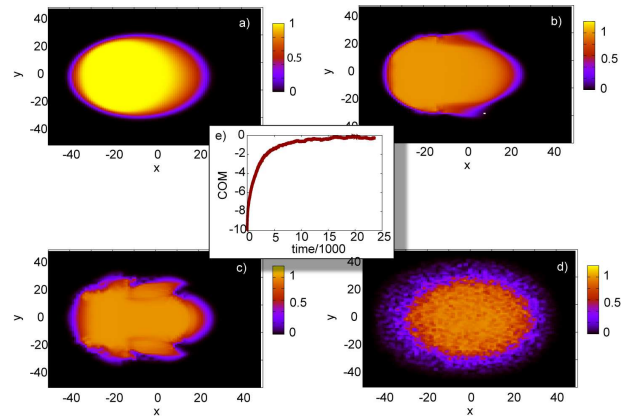}
\vspace{-.2cm}
\caption{(Color online) Density distributions for a system of strongly interacting bosons after a shift in the harmonic potential on a square lattice with $V_0=0.016$, $A=10$ and $N=2551$ and $U=40$ For these parameters, a very large Mott plateau forms in the center. Still the center of mass fully relaxes to zero as shown in e). 
The four panels display the density profile for a) $t/J=15$,  b) $t/J=30$, c) $t/J=45$ and d) $t/J=9000$. 
}
\label{2dmott}
\end{figure}

\begin{figure}
\includegraphics[scale=0.31]{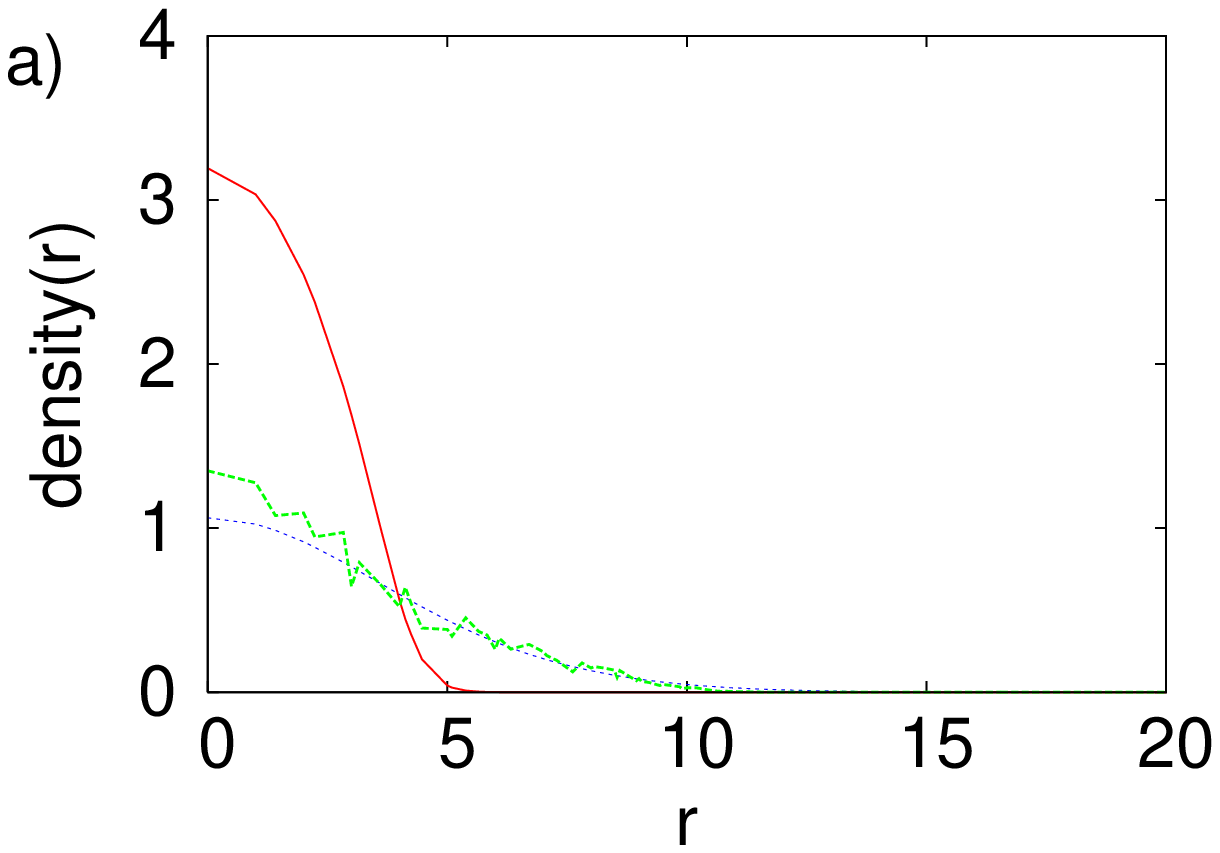}
\includegraphics[scale=0.31]{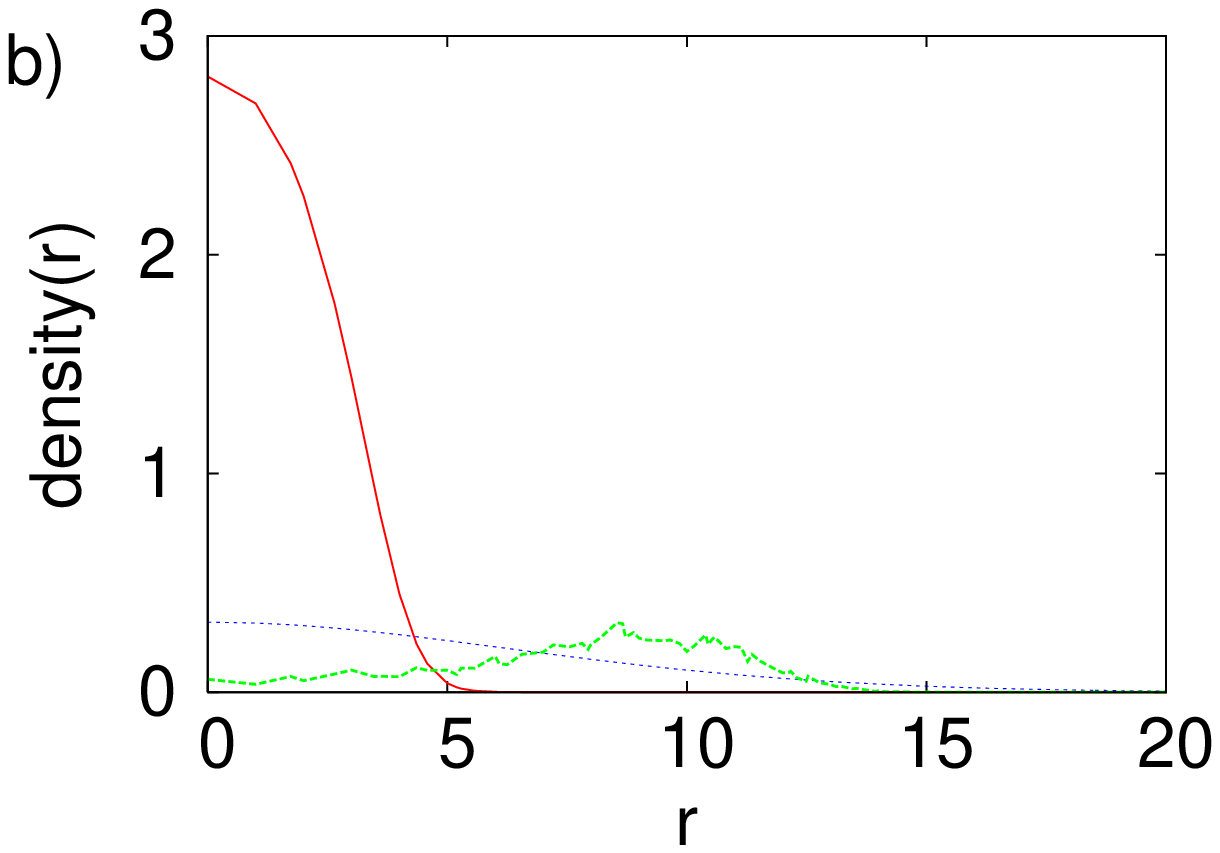}
\includegraphics[scale=0.31]{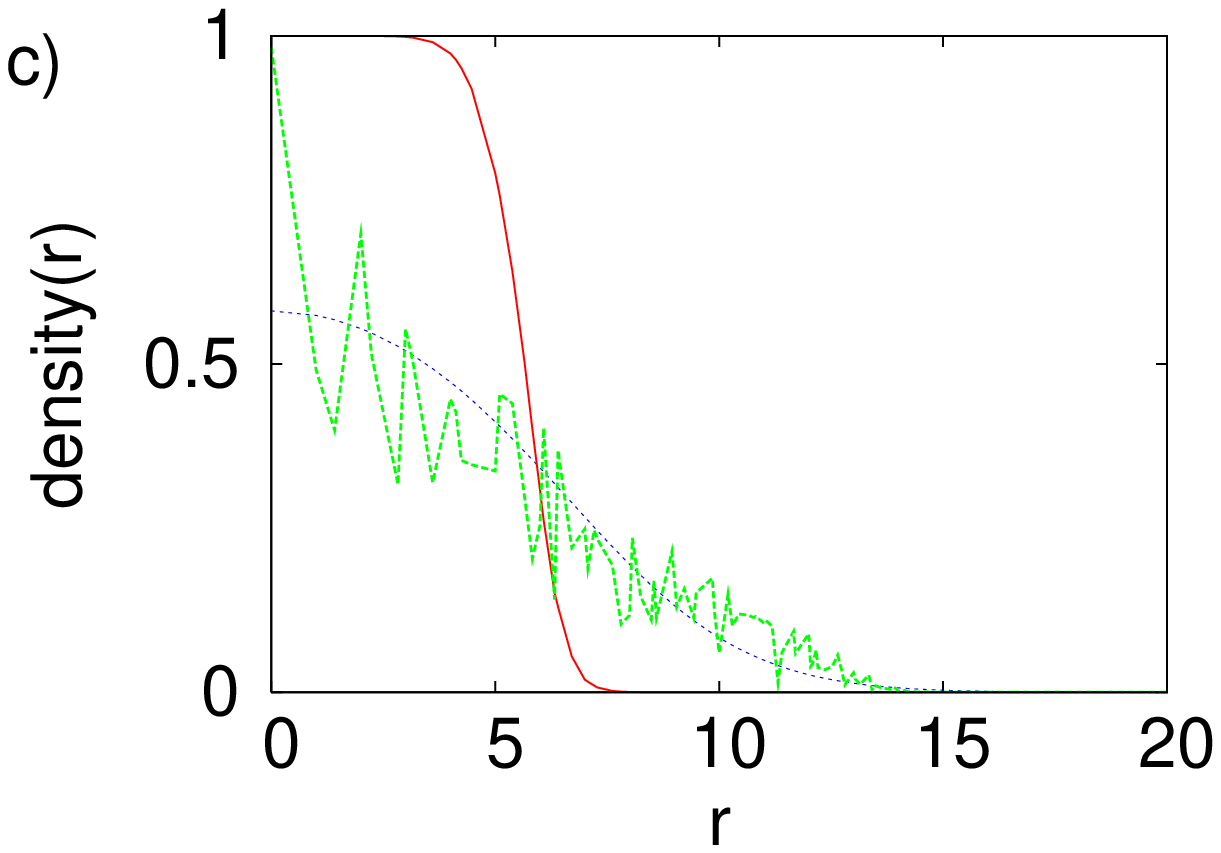}
\includegraphics[scale=0.31]{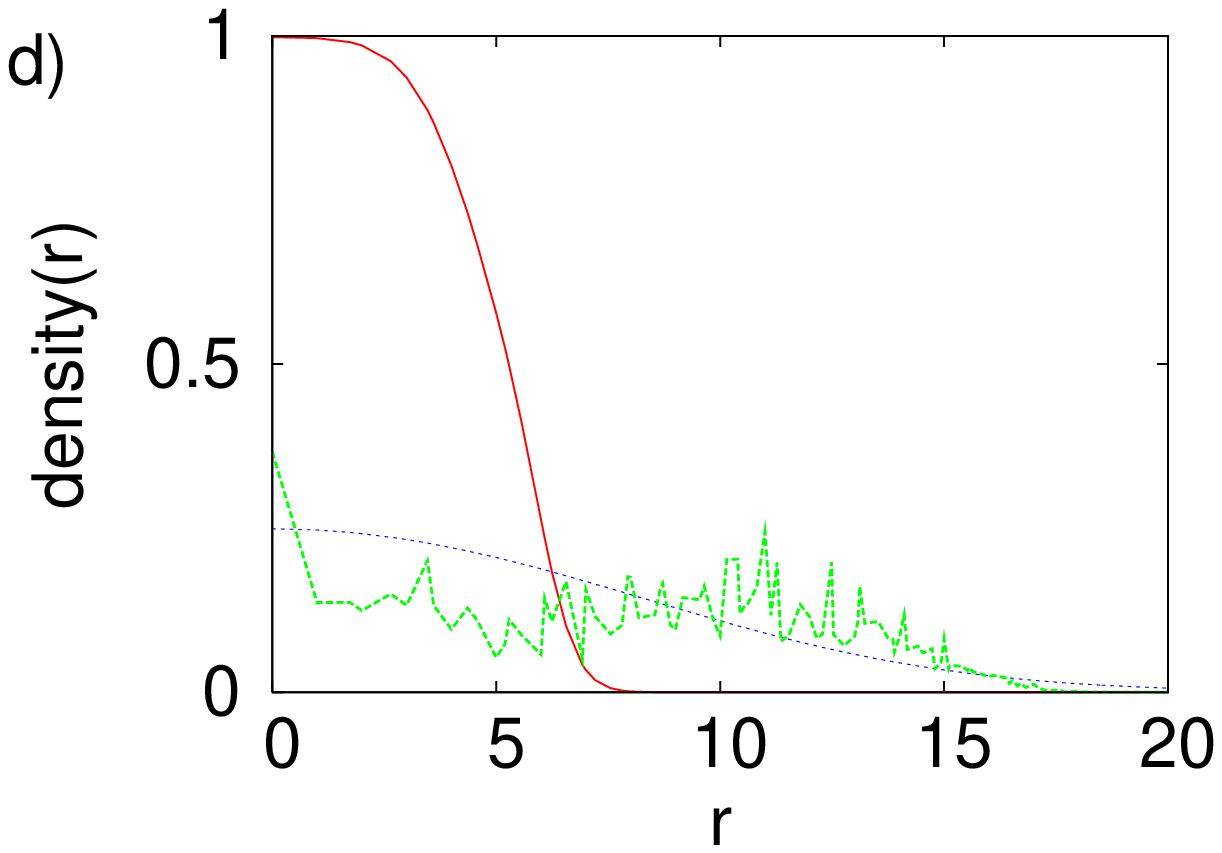}
\vspace{-0.3cm}
\caption{(Color online) Radial density profile before the shift of the harmonic potential (red) and after relaxation to the new equilibrium (green) for the square lattice (left pictures) and hexagonal lattice (right pictures) for $U=2$ (upper row) and $U= \infty$ (lower row). The blue line is a thermal density profile. The bosons on the square lattice have relaxed from a shift of six lattice sites; the bosons on the hexagonal lattice relaxed from a shift of ten lattice sites. Other parameters are chosen as $N=100$ and $V=0.3$.}
\label{raddens}
\end{figure}

We now turn to strongly interacting bosons, for which the dimensionality of the lattice is very important. In one spatial dimension, the motion gets completely blocked when a Mott plateau is present, because the atoms can not pass each other and the center of mass never reaches the new equilibrium position \cite{Rigol05}. This behavior is reproduced by our Gutzwiller calculations as shown in Fig. \ref{1dmott}. In two dimensions, however, the center of mass motion shows dissipative dynamics and fully relaxes to the new equilibrium (Fig. \ref{2dcom}). We did extensive calculations for a wide range of parameters and always found this behavior. The reason is that the superfluid shell can freely move around the Mott-insulating core, thus allowing the system to relax. This is visible in the density plots in \mbox{Fig. \ref{2dmott}}, which show that initially the Mott plateau remains inert, but afterwards shows dynamical melting induced by the dynamics of the superfluid ring. Moreover, the insulating plateau scatters the atoms away from their single-particle trajectories, which leads to a significant broadening of the density profile. Therefore, the final state is not the equilibrium state in the shifted potential, because the total energy is conserved. In particular, the Mott plateau has disappeared (Fig. \ref{2dmott}d)). The suppressed density in the center and the  broadening of the density profile is shown in Fig. \ref{raddens}. We compare this with a finite temperature density profile, which is calculated neglecting phase fluctuations. We derive the temperature from the energy induced in the system by the shift in the potential. This leads to a good agreement, which shows that the density profiles for those parameters thermalize. 
The complete relaxation of the center of mass even persists for hard-core bosons.  
It is worth noticing that the time-scale for the relaxation cannot easily be expressed in terms of the interaction strength, because the relaxation is not due to single-particle tunneling processes. Therefore, the observed relaxation of the center of mass is very slow. This is partly due to the parameters chosen for the simulations; taking a  more shallow harmonic potential and a larger number of particles, as experimentally more realistic, brings the timescale back into the experimentally relevant regime.
On the hexagonal lattice, the qualitative behavior of strongly interacting bosons is the same as above: the center of mass always relaxes to zero. However, quantitatively the lattice topology is very important: the dynamics on the hexagonal lattice is  much faster (Fig. \ref{2dcom}b) and d)) and equal to the single-particle relaxation time. This is because the single particle dynamics is along the equipotential lines. Therefore the atoms avoid each other and are able to relax faster. From Fig. \ref{raddens} b) and d) it is obvious that this leads to a density-profile that is suppressed in the center and has the same ring-like structure as the non-interacting bosons obtain after relaxation of the center of mass. As an important consequence, this implies that the density profile does not thermalize on the hexagonal lattice, because the thermal density profile is always maximal in the center of the trap.

In conclusion, we studied the effect of strong interactions, dimensionality and lattice topology on the particle transport of interacting bosons induced by a change in the underlying harmonic potential. We find that on the square lattice the single-particle dynamics on the square lattice is insulating after a large shift, and can be described in terms of Bloch oscillations. On the hexagonal lattice the single particle dynamics shows relaxation along the equipotential lines. For strongly interacting bosons, we find that in two dimensions the presence of a Mott plateau does not prohibit relaxation of the center of mass to the new equilibrium, in contrast to the one-dimensional situation. This relaxation is very slow on the square lattice, but much faster on the hexagonal lattice.  Our numerical predictions can be directly verified experimentally by performing absorption measurements, from which both the center of mass dynamics and the density profiles can be obtained.

\vspace{-.05cm}

We thank Michael K\"ohl and Henning Moritz for useful discussions. This work was supported by the SFB-TRR 49 of the German Science Foundation (DPG).


\vspace{-.6cm}

\begin{thebibliography}{99}

\bibitem{Jaksch98} D. Jaksch {\it et al.}, Phys. Rev. Lett. {\bf 81}, 3108 (1998).

\bibitem{Greiner02} M. Greiner {\it et al.},  Nature {\bf 415}, 39 (2002).

\bibitem{Hofstetter02} W. Hofstetter {\it et al.}, Phys. Rev. Lett. {\bf 89}, 220407 (2002).

\bibitem{Kohl05} M. K\"ohl {\it et al.}, Phys. Rev. Lett. {\bf 94}, 080403 (2005).

\bibitem{Bloch07} I. Bloch, J. Dalibard and W. Zwerger, arXiv:0704.3011.

\bibitem{Morsch06} O. Morsch and M. Oberthaler, Rev. Mod. Phys. {\bf 78}, 179 (2006).

\bibitem{Stoferle04} T. St\"oferle {\it et al.}, Phys. Rev. Lett. {\bf 92}, 130403 (2004).

\bibitem{Kollath06} C. Kollath {\it et al.}, Phys. Rev. Lett. {\bf 97}, 050402 (2006). 

\bibitem{Cataliotti01} F.S. Cataliotti {\it et al.}, Science {\bf 293}, 843 (2001).

\bibitem{Burger01} S. Burger {\it et al.}, Phys. Rev. Lett. {\bf 86}, 4447 (2001).

\bibitem{Cataliotti03} F.S. Cataliotti {\it et al.}, New J. Phys. {\bf 5}, 71 (2003).

\bibitem{Tuchman06} A.K. Tuchman {\it et al}., N. J. Phys. {\bf 8}, 311 (2006).

\bibitem{Fertig05} C.D. Fertig {\it et al}., Phys. Rev. Lett {\bf 94}, 120403 (2005).

\bibitem{Henderson06} K. Henderson {\it et al.}, Phys. Rev. Lett {\bf 96}, 150401 (2006). 

\bibitem{Polkovnikov04} A. Polkovnikov and D.-W. Wang, Phys. Rev. Lett. {\bf 93}, 070401 (2004).

\bibitem{Rigol05} M. Rigol {\it et al}., Phys. Rev. Lett. {\bf 95}, 110402 (2005).

\bibitem{Ruostekoski05} J. Ruostekoski and L. Isella, Phys. Rev. Lett. {\bf 95}, 110403 (2005).

\bibitem{Rey05} A. M. Rey {\it et al.}, Phys. Rev. A {\bf 72}, 033616 (2005).

\bibitem{Pupillo06} G. Pupillo, A. M. Rey, C. J. Williams, C. W. Clark, New J. Phys. {\bf 8}, 161 (2006).

\bibitem{Gea06} J. Gea-Banacloche {\it et al.}, Phys. Rev. A {\bf 73}, 013605 (2006).

\bibitem{Mudugno03} G. Modugno {\it et al}., Phys. Rev. A {\bf 68}, 011601(R) (2003).

\bibitem{Pezze04}  L. Pezz\`e {\it et al}., Phys. Rev. Lett. {\bf 93}, 120401 (2004). 

\bibitem{Altman05}
 E. Altman {\it et al.}, Phys. Rev. Lett. {\bf 95}, 020402 (2005).

\bibitem{Polkovnikov05}
 A. Polkovnikov {\it et al}., Phys. Rev. A {\bf 71}, 063613 (2005).

\bibitem{Mun07} J. Mun {\it et al.}, arXiv:0706.3946.

\bibitem{Zuerich} This work was motivated by ongoing experiments on the dynamics of fermionic atoms in a three-dimensional optical lattice at ETH Z\"urich (private communication).


\bibitem{Jaksch02} D. Jaksch {\it et al}., Phys. Rev. Lett. {\bf 89}, 040402 (2002).

\bibitem{Buchleitner03} Andreas Buchleitner and Andrey R. Kolovsky, Phys. Rev. Lett. {\bf 91}, 253002 (2003).

\bibitem{Hooley04} C. Hooley and J. Quintanilla, Phys. Rev. Lett. {\bf 93}, 080404 (2004).

\bibitem{Ponomarev05} A. V. Ponomarev and A. R. Kolovsky, Las. Phys. {\bf 16}, 367 (2006).


\end{thebibliography}
\end{document}